\documentclass[12pt]{iopart}
\usepackage{amssymb}
\usepackage{amsfonts}
\usepackage{amsgen}
\usepackage{amsbsy}
\usepackage{setstack}

\newcommand{\beq}{\begin{equation}}
\newcommand{\eeq}{\end{equation}}

\newcommand{\bea}{\begin{eqnarray}}
\newcommand{\eea}{\end{eqnarray}}

\begin{document}

\title{Optimal unbiased state characterization }
\author{ I Sainz, J~J~D\'{\i}az and A~B~Klimov}

\begin{abstract}
We propose a general approach to characterize states of a bipartite system
composed by a fully controllable and an unaccessible subsystems. The method
is based on the measuring interference between states of the uncontrollable
subsystem obtained after projecting an appropriately transformed bipartite
state on the basis of the accessible subsystem by local operations.
\end{abstract}

\maketitle

\address{Departamento de F\'{i}sica, Universidad de Guadalajara, Revoluci\'{o}n
1500, Guadalajara, Jalisco 44420, M\'{e}xico }

\section{Introduction}

The characterization of quantum states is a central problem in quantum
information science. The full information about a state with no or little
prior information can be obtained by quantum tomography (QT) protocols \cite%
{QT,QT1,QT2,QT3,QT4,QT5,QT6}. These are usually very resource demanding
tasks, since numerous copies of the original state are required in order to
accumulate reliable statistics about measured probabilities \cite%
{Blatt,Blatt1,Blatt2}. Different strategies for QT have been theoretically
proposed and experimentally implemented \cite%
{QTexp,QTexp1,QTexp2,QTexp3,QTexp4,QTexp5}. Nevertheless, even after a large
set of measurements the experimental results still should be
\textquotedblleft cleaned\textquotedblright\ by applying sophisticated
mathematical methods \cite{max likelihood,max likelihood1,max likelihood2}
in order to insure the best estimation of the reconstructed state.

On the other hand the quantum state discrimination (QSD) problem \cite%
{QSD1a,QSD1b,QSD1c} (for recent reviews see \cite{QSD2a,QSD2b,QSD2c} and
references therein), essential in numerous quantum information protocols, is
placed. In this case the issue is to determine in what state, chosen from a
set of known states, was the system prepared. This problem is also far from
being simple especially if one is required to find an optimal way for a
conclusive and unambiguous state discrimination. The principal challenge
here is to find an experimentally feasible set of POVMs, the construction of
which usually requires to enlarge the dimension of the quantum system by
adding an adequate auxiliary system (ancilla). The ancillary system should
be easily accessible and completely manageable so that all desirable
operations can be implemented, and in particular, a combination of unitary
operations in the Hilbert space of bi-partite (or in general multi-partite)
systems plus von Neumann projections allow to carry on QSD protocols.

A general bipartite state\ (system + ancila) can be represented in form of
the Schmidt decomposition

\begin{equation}
\left\vert \Psi \right\rangle =\sum_{l=1}^{d}\sqrt{\lambda _{l}}\left\vert
\psi _{l}\right\rangle _{s}\left\vert u_{l}\right\rangle _{a},  \label{SD}
\end{equation}%
where $\lambda _{l}$ are positive numbers, $d$ is the so called Schmidt rank
and subscripts $s$ and $a$ are correspondingly for system and ancilla
states. Even in the case when we do not have access to the system , i.e. we
are not able to apply any operations to the system's Hilbert space, it is
still possible to determine basic characteristics of the elements of the
Schmidt decomposition (\ref{SD}). We can just project out the bipartite
state into the ancillary subspace, obtaining a reduced density matrix
\begin{equation}
\rho _{a}=Tr_{s}\left\vert \Psi \right\rangle \langle \Psi |=\sum_{k,n}\mu
_{kn}\left\vert u_{k}\right\rangle \langle u_{n}|,  \label{rhoa}
\end{equation}%
where the coefficients $\mu _{kn}$ depend on the parameters $\lambda _{l}$
and the scalar products $\lambda _{l^{\prime }l}=\left\langle \lambda
_{l^{\prime }}|\lambda _{l}\right\rangle $. Since it is supposed that the
ancilla is completely controllable, the density matrix $\rho _{a}$ can be
reconstructed by applying standard tomographic methods, and thus,
determining the above mentioned unknown parameters. Nevertheless, as we
mentioned before, this is a very \textquotedblleft
expensive\textquotedblright\ way to characterize the decomposition (\ref{SD}%
). In this paper we propose an alternative way to determine the parameters $%
\{\lambda _{l},\lambda _{l^{\prime }l},l,l^{\prime }=1,...,d\}$. When the
Schmidt rank $d$ is a prime number it is possible to combine the state
discrimination protocols and the tomography methods based on the mutually
unbiased bases (MUB) \cite{MUBtom,MUBtom1,MUBtom2,MUBtom3,MUBtom4,MUBtom5}.
The main idea consists in applying specific unitary transformations (
exactly those that are used for MUBs generation in prime dimensions) to the
ancillary system with a consecutive projection on a basis in the Hilbert
space of the ancilla. In this way one can obtain a set of system states such
that all the required information about the state (\ref{SD}) can be obtained
from the interference picture (mutual projections of the resulted system
states). In other words, we can obtain exactly the same information as in
tomographic reconstruction of the ancillary density matrix (\ref{rhoa}), but
using less resources.

The paper is organized as follows: In Sec.II we describe the general method,
applicable in any prime dimension and complex scalar products $\lambda
_{l^{\prime }l}$. In Sec.III we provide a method of solving the equations
derived in Sec.II in the particular case of real $\lambda _{l^{\prime }l}$.
In Sec.IV we give explicit solutions in two and three dimensional cases.

\section{The method}

Let us consider the Schmidt decomposition of a bipartite quantum system
composed by an unknown and uncontrollable subsystem $\Lambda $ and a
controllable subsystem $U$,
\begin{equation}
\left\vert \Psi _{0}\right\rangle =\sum_{l=0}^{p-1}\sqrt{\lambda _{l}}%
\left\vert \lambda _{l}\right\rangle \left\vert u_{l}\right\rangle ,
\label{Psi0}
\end{equation}%
where $\{\left\vert \lambda _{l}\right\rangle \}$ and $\{\left\vert
u_{l}\right\rangle \}$ are states of subsystems $\Lambda $ and $U$
respectively, and the Schmidt rank is a prime number $p$. In practice,
bipartite states of this form are obtained after applying to a factorized
state a combination of local and conditional unitary transformations \cite%
{mataloni}. The set $\{\left\vert u_{l}\right\rangle \}$ can be chosen
orthonormal, $\langle u_{l}|u_{k}\rangle =\delta _{lk}$, leading to the
normalization condition on the \textquotedblleft
probabilities\textquotedblright\
\begin{equation}
\sum_{l=0}^{p-1}\lambda _{l}=1.  \label{norm}
\end{equation}

In order to extract the maximum possible information, i.e. the probabilities
$\lambda _{l}$ and the scalar products $\lambda _{l^{\prime }l}=\left\langle
\lambda _{l^{\prime }}|\lambda _{l}\right\rangle $, about the state $%
\left\vert \Psi _{0}\right\rangle $ we apply to the system $U$ the whole set
of Hadamard transformations $H_{s}$. In prime dimensions this set consists
in $p$ transformations of the form
\begin{equation}
H_{s}=H_{0}D^{s}=\frac{1}{\sqrt{p}}\sum_{l,k=0}^{p-1}\omega
^{-2^{-1}sl^{2}-kl},\quad s=0,\ldots ,p-1  \label{HD}
\end{equation}%
where $\omega =e^{2i\pi /p}$ is the $p$th root of unity, and $e^{-i\phi
}=\omega ^{-2^{-1}}$ for $p>2$, while $e^{-i\phi }=-i$ for $p=2$. Here, all
the operations in the exponent of $\omega $ ($e^{-i\phi }$) are modulus $p$.
The set of Hadamard matrix (\ref{HD}) is closely related to the standard
form of MUB construction \cite{HD,HD1}. Really, the columns of matrices $%
H_{s}$ are elements of bases which are unbiased for different values of the
index $s$\emph{.} In this Section we focus on the case $p>2$. The dimension $%
p=2$ will be studied separately in Section IV.
\begin{equation}
H_{0}=\frac{1}{\sqrt{p}}\sum_{l,k=0}^{p-1}\omega ^{-lk}\left\vert
u_{l}\right\rangle \left\langle u_{k}\right\vert ,  \nonumber
\end{equation}%
is the finite Fourier transform operator and $D$ is the diagonal operator
\begin{equation}
D=\sum_{l=0}^{p-1}e^{-i\phi l^{2}}\left\vert u_{l}\right\rangle \left\langle
u_{l}\right\vert ,  \nonumber
\end{equation}

After applying the transformation $I^{(\Lambda )}\otimes H_{s}^{(U)}$ to the
state (\ref{Psi0}) and projecting over every state $\left\vert
u_{k}\right\rangle $ of the controllable subsystem we obtain the following
normalized $p$ states of the system $\Lambda $:
\begin{equation}
\left\vert \psi _{k}^{s}\right\rangle =\frac{1}{\sqrt{N_{ks}}}%
\sum_{l=0}^{p-1}\omega ^{-2^{-1}sl^{2}-kl}\sqrt{\lambda _{l}}\left\vert
\lambda _{l}\right\rangle ,\quad k=0,\ldots ,p-1.  \label{psiks}
\end{equation}%
The normalization factors are given by
\begin{equation}
N_{ks}=1+\sum_{ l,l^{\prime }=0\;l^{\prime }\neq
l}^{p-1}\omega^{2^{-1}s(l^{\prime 2}-l^{2})+k(l^{\prime }-l)}x_{l^{\prime
}l},  \label{Nks}
\end{equation}
where we have introduced $x_{l^{\prime }l}=\lambda _{l^{\prime }l}\sqrt{%
\lambda _{l^{\prime }}\lambda _{l}}$, note that these factors automatically
fulfill the relations
\begin{equation}
\sum_{k=0}^{p-1}N_{ks}=p,  \label{relNks}
\end{equation}%
for every $s=0,\ldots ,p-1$.

The state (\ref{Psi0}) contains $p^{2}$ unknown parameters: $p$ real
(positive) probabilities $\lambda _{l}$, and $p(p-1)/2$ complex inner
products $\lambda _{l^{\prime }l}$. These parameters can be determined by
measuring projections (obtained from interference experiments) of (\ref%
{psiks}) with $s=0,..,p-1$ into a single state $\left\vert \psi
_{0}^{0}\right\rangle $. Then, we have a set of (complex) equations of the
form
\begin{equation}
\langle \psi _{0}^{0}\left\vert \psi _{k}^{s}\right\rangle =a_{ks},
\label{Meq}
\end{equation}%
where the right hand side are measurable quantities. Explicitly, for a given
$s=1,\ldots ,p-1$ we obtain a set of $p$ equations,
\begin{equation}
a_{ks}\sqrt{N_{00}N_{ks}}=\sum_{l=0}^{p-1}\omega ^{-2^{-1}sl^{2}-kl}\lambda
_{l}+\sum_{ l,l^{\prime \prime }}^{p-1}\omega
^{-2^{-1}sl^{2}-kl}x_{l^{\prime }l},  \label{ecks}
\end{equation}%
while for $s=0$ one has $p-1$ equations of the form
\begin{equation}
a_{k0}\sqrt{N_{00}N_{k0}}=\sum_{l=0}^{p-1}\omega ^{-kl}\lambda _{l}+\sum _{
l,l^{\prime \prime }}^{p-1}\omega ^{-kl}x_{l^{\prime }l},  \label{eck0}
\end{equation}%
where $k=1,\ldots ,p-1$.

The set of equations (\ref{ecks}) contains redundant information about the
system. In particular, the coefficients of $x_{l^{\prime }l}$ and $\lambda
_{l}$ in Eqs. (\ref{ecks}) labeled by $p-s$ and $p-k$ are complex conjugated
to those labeled by $s$ and $k$\textbf{,} leading to the same complex
equation when the inner products are real. Thus, the information that can be
extracted from Eq. (\ref{ecks}) with $s=(p+1)/2,...p-1$ is just the same as
the one we can obtain from the first $(p-1)/2$ sets. The same situation
holds for the set corresponding to $s=0$: only the first $(p-1)/2$ equations
in (\ref{eck0}) provide non-redundant information about the system. These
leaves us with $p(p-1)/2+(p-1)/2$ complex equations, which together with the
normalization condition (\ref{norm}) gives exactly $p^{2}$ independent real
equations.

The equations (\ref{ecks}) and (\ref{eck0}) are nonlinear, which is a clear
drawback in comparison to the standard unbiased tomographic scheme.
Nevertheless, in the special case of real scalar products $\lambda
_{l^{\prime }l}$, these sets of equations can be quasilinearized, and thus,
an analytic solution for an arbitrary dimension $p$ can be found.

\section{Real inner products}

In the case when $\lambda _{l^{\prime }l}$ are real numbers the number of
the real parameters to determine is reduced to $p(p+1)/2$.

It is convenient to rewrite Eqs.(\ref{ecks}) in the form
\begin{equation}
\sqrt{N_{00}}\omega ^{2^{-1}sl^{2}}\sum_{k=0}^{p-1}\omega
^{kl}a_{ks}z_{ks}=py_{l},\quad l=0,\ldots ,p-1,  \label{eclin}
\end{equation}%
where we have introduced the new real variables
\begin{equation}
y_{l}=\lambda _{l}+\sum_{ l^{\prime }=0 \;l^{\prime }\neq l}^{p-1}
x_{l^{\prime }l},  \label{yl}
\end{equation}%
and $z_{ks}=\sqrt{N_{ks}}=\sqrt{N_{p-ks-k}}$, for $s=1,\ldots
,(p-1)/2$. The variables $z_{ks}$ are not independent since the relation (%
\ref{relNks}) imposes the following restrictions
\begin{equation}
\sum_{k=0}^{p-1}z_{ks}^{2}=p,  \label{norm z}
\end{equation}%
for any $s$. The normalization factor $N_{00}$ is related to the new
variables through the relation%
\begin{equation}
N_{00}=\sum_{l=0}^{p-1}y_{l}=\frac{1}{p^{2}}\left[
\sum_{k=0}^{p-1}a_{ks}z_{ks}G(2^{-1}s,k)\right] ^{2},  \label{N00}
\end{equation}%
where $G(s,k)=\sum_{l=0}^{p-1}\omega ^{sl^{2}}\omega ^{kl}$ is the Gauss
sum. The last equality in (\ref{N00}), obtained by summing up equations in (%
\ref{eclin}), holds \ for any $s=1,\ldots ,(p-1)/2$, and implies that the
measured quantities $a_{ks}$ are not independent if the scalar products $%
\lambda _{l^{\prime }l}$ are real. Indeed, it is sufficient to measure $p-1$
imaginary parts of $\langle \psi _{0}^{0}\left\vert \psi
_{k}^{s}\right\rangle $, for $s=1,...,(p-1)/2$ and the real part for only a\
single (arbitrary) value of $s$. For the complete characterization of the
state (\ref{Psi0}) we need another $(p-3)/2$ measurements from $s=0$.

The condition that the imaginary part of (\ref{eclin}) is zero leads to $p-1$
linearly independent equations%
\begin{equation}
\mathrm{Im}\left( \omega ^{2^{-1}sl^{2}}\sum_{k=0}^{p-1}\omega
^{kl}a_{ks}z_{ks}\right) =0,\quad l=0,\ldots ,p-1,  \label{Ims}
\end{equation}%
for a given value of $s=1,\ldots ,(p-1)/2$, which together with the
condition (\ref{norm z}) allow to determine all $z_{ks}$, for $s=1,\ldots
,(p-1)/2$, and thus all $N_{ks}$, $s=1,\ldots ,p-1$ since for real $\lambda
_{l^{\prime }l}$ we have $N_{ks}=N_{p-k\,s-k}$. Having obtained $z_{ks}$,
the variables $y_{l}$ are immediately determined from the real part of (\ref%
{eclin}) for any $s$ and (\ref{N00}).

For $s=0$ Eq. (\ref{eclin}) is reduced to the following form,
\begin{equation}
\sqrt{N_{00}}\sum_{k=0}^{p-1}\omega ^{kl}a_{k0}z_{k0}=py_{l},\quad
l=1,...,(p-1)/2,  \label{eclin0}
\end{equation}%
where $a_{00}=\langle \psi _{0}^{0}|\psi _{0}^{0}\rangle =1$, $%
a_{p-k0}=a_{k0}^{\ast }$, and the variables $z_{k0}$ satisfy the symmetry
condition $z_{p-k0}=z_{k0}$. Equation (\ref{relNks}) now reads
\begin{equation}
N_{00}+2\sum_{k=1}^{\frac{p-1}{2}}z_{k0}^{2}=p.  \label{N000}
\end{equation}%
As above, the variables $z_{k0},k=1,...,(p-1)/2$ are obtained form the
conditions%
\begin{equation}
\mathrm{Im} \left( \sum_{k=1}^{\frac{p-1}{2}}\omega ^{kl}a_{k0}z_{k0}\right)
=0,\quad l=1,...,(p-1)/2,  \label{Im0}
\end{equation}%
together with (\ref{N000}).

Finally, having obtained all $N_{ks}$ and $y_{l}$ we can invert Eqs. (\ref%
{Nks}) and (\ref{yl}) to determine the physical parameters $\lambda _{l}$
and $\lambda _{l^{\prime }l}$. It should be stressed here that since $N_{ks}$
are not linearly independent for a fixed value of the index $s=0,...,p-1$,
as it follows from Eqs. (\ref{Ims}), (\ref{Im0}), there are $(p-1)^2/2-1$
linearly independent $N_{ks}$ and $p$ linearly independent $y_{l}$, so that
the normalization condition (\ref{norm}) should be added in order to be able
to reconstruct $\lambda _{l}$ and $\lambda _{l^{\prime }l}$.

In the next Section we show how this approach works in the particular cases
of $p=2$ and $p=3$.

\section{Examples}

\subsection{Dimension two}

In the case $p=2$ we have $\omega =-1$ and $\omega ^{-2^{-1}}\rightarrow -i$%
, so that equations (\ref{ecks}) read as
\begin{eqnarray}
\sqrt{N_{00}N_{01}}a_{01} &=&y_{0}+iy_{1},  \label{2.2a} \\
\sqrt{N_{00}N_{11}}a_{11} &=&y_{0}-iy_{1},  \label{2.2b}
\end{eqnarray}%
where $y_{0}=\lambda _{0}+x_{10},~y_{1}=\lambda _{1}+x_{01}$, and
\[
N_{00}=1+2x_{R},\qquad N_{01}=1-2x_{I},\qquad N_{11}=1+2x_{I},
\]%
here we have introduced $x_{01}=x_{R}+ix_{I}$. Observe that similarly to the
case $p>2$, Eqs. (\ref{2.2a}) and (\ref{2.2b}) provide the same information,
so that we pick for instance Eq. (\ref{2.2a}). From (\ref{eck0}) we obtain a
single relation
\begin{equation}
\sqrt{N_{00}N_{10}}a_{10}=y_{0}-y_{1},  \label{2.1}
\end{equation}%
with
\[
N_{01}=1-2x_{R},
\]%
where again, only the imaginary part will be considered. Introducing
imaginary and real parts of the measurement as $a_{10}=\alpha _{0}+i\beta
_{0}$, and $a_{01}=\alpha _{1}+i\beta _{1}$ we arrive to three real
equations
\begin{eqnarray}
\sqrt{(1+2x_{R})(1-2x_{I})}\alpha _{1}& =&\lambda _{0}+x_{R}-x_{I},
\label{2} \\
\sqrt{(1+2x_{R})(1-2x_{I})}\beta _{1} &=&\lambda _{1}+x_{R}-x_{I},  \label{3}
\end{eqnarray}%
\begin{equation}
\sqrt{1-4x_{R}^{2}}\beta _{0} =-2x_{I},  \label{1}
\end{equation}
where the first two equations corresponds to the real and the imaginary part
of (\ref{2.2a}), while the last one is the imaginary part of (\ref{2.1}).
Eqs. (\ref{2})-(\ref{1}) together with the normalization condition $\lambda
_{0}+\lambda _{1}=1$ allow to determine four real parameters, $\lambda
_{0,1} $ and $x_{R,I}=\sqrt{\lambda _{0}\lambda _{1}}\mathrm{Re}(\mathrm{Im}%
)\lambda _{01}$.

\subsection{Dimension three}

Here we discuss the case of $p=3$ when the scalar products between any pair
of states $\left\vert \lambda _{l}\right\rangle $ are real.

Eqs. (\ref{eclin}) for $s=1$ take the form
\begin{eqnarray}
\sqrt{N_{00}}\left( z_{01}a_{01}+z_{11}a_{11}+z_{21}a_{21}\right)& =&3y_{0}
\label{3.1} \\
\sqrt{N_{00}}\left( \omega ^{2}z_{01}a_{01}+z_{11}a_{11}+\omega
z_{21}a_{21}\right) &=&3y_{1}  \label{3.2} \\
\sqrt{N_{00}}\left( \omega ^{2}z_{01}a_{01}+\omega
z_{11}a_{11}+z_{21}a_{21}\right) &=&3y_{2},  \label{3.3}
\end{eqnarray}%
where $\omega =e^{2\pi i/3}$,
\[
z_{11}^{2}=N_{11}=1+2x_{01}-x_{02}-x_{21},\quad
z_{21}^{2}=N_{21}=1-x_{01}+2x_{02}-x_{21},
\]%
and $z_{01}^{2}=N_{01}=3-N_{11}-N_{21}$.

Following the procedure of Sec. III, we express $z_{11}$ and $z_{21}$ as
functions of $z_{01}$ form the imaginary part of Eqs. (\ref{3.2}), (\ref{3.3}%
), and then using (\ref{relNks}) one obtains $z_{01}$ in terms of the
measured quantities,
\begin{eqnarray}
z_{01} &=&\frac{\sqrt{3}(\alpha _{1}\alpha _{2}-3\beta _{1}\beta _{2})}{R},
\label{z1} \\
z_{11} &=&\frac{\sqrt{3}\alpha _{0}(\alpha _{2}-\sqrt{3}\beta _{2})}{R},
\label{z2} \\
z_{21} &=&\frac{\sqrt{3}\alpha _{0}(\alpha _{1}-\sqrt{3}\beta _{1})}{R},
\label{z3}
\end{eqnarray}%
where
\[
R=\left( \left( \alpha _{1}\alpha _{2}-3\beta _{1}\beta _{2}\right)
^{2}+\alpha _{0}^{2}\left[ \left( \alpha _{1}-\sqrt{3}\beta _{1}\right)
^{2}+\left( \alpha _{2}-\sqrt{3}\beta _{2}\right) ^{2}\right] \right)
^{1/2},
\]%
and we have introduced $a_{j1}=\alpha _{j}+i\beta _{j}$, $j=0,1,2$. It worth
noting that $\beta _{0}$ does not appear in the above expressions. Now,
following the general procedure we obtain
\begin{equation}
N_{00}=\frac{1}{3}\left[ 2\beta _{0}z_{01}+\left( \sqrt{3}\alpha _{1}-\beta
_{1}\right) z_{11}+\left( \sqrt{3}\alpha _{2}-\beta _{2}\right) z_{21}\right]
^{2},  \label{N003}
\end{equation}%
where $z_{j1}$ are given in (\ref{z1})-(\ref{z3}).

Using (\ref{z1})-(\ref{z3}) and (\ref{N003}) it is straightforward to
express $y_{l}$ as a function of $z_{j1}$ and $N_{j1},N_{00}$ form the real
part of Eqs. (\ref{3.1})-(\ref{3.3}). It is worth noticing here that $N_{10}$
can be found directly from $2N_{10}=3-N_{00}$. Now we can completely
characterize the initial state by the probabilities
\begin{eqnarray}
\lambda _{0} &=&y_{0}+\frac{1}{3}\left( N_{01}-N_{00}\right) ,  \nonumber \\
\lambda _{1} &=&y_{1}+\frac{1}{3}\left( N_{21}-N_{00}\right) ,  \nonumber \\
\lambda _{2} &=&y_{2}+\frac{1}{3}\left( N_{11}-N_{00}\right) ,  \nonumber
\end{eqnarray}%
and the inner products $\lambda _{l^{\prime }l}=x_{l^{\prime }l}/\sqrt{%
\lambda _{l}^{\prime }\lambda _{l}}$, where
\begin{eqnarray}
x_{01} &=&\frac{1}{6}\left( N_{00}+N_{11}-N_{01}-N_{21}\right) ,  \nonumber
\\
x_{02} &=&\frac{1}{6}\left( N_{00}+N_{21}-N_{01}-N_{11}\right) ,  \nonumber
\\
x_{12} &=&\frac{1}{6}\left( N_{00}+N_{01}-N_{11}-N_{21}\right) .  \nonumber
\end{eqnarray}

\bigskip

\section{Conclusions}

We proposed a general characterization of bipartite states of the form (\ref%
{SD}) in the case when one of the subsystems is not accessible by unitary
transformations. We have shown that such characterization can be done
without applying the complete tomographic procedure but only measuring
interference between states of the uncontrollable subsystem obtained after
projecting an appropriately transformed bipartite state on the basis of the
accessible (by local operations) subsystem.

In this approach we can reduce the number of required quantum resources and
also avoid to process a vast amount of statistical information usually
obtained from tomographic data. In fact we only need $p^2-1$ copies of the
initial state required to produce the $(p^2-1)/2$ projections to state $%
\left\vert \psi_0^0 \right\rangle$ that we need to characterize the desired
state. In this sense the characterization we proposed is optimal since we
require the same number of measurements (and not only setups, as in the MUB
tomography scheme) as the number of unknown parameters.

In the present approach we have used the properties of the Hadamard
transformations in the case of prime dimension, which allows us to approach
to the analytical solution of the general problem. These Hadamard
transformations are exactly those that generates a complete set of mutually
unbiased bases. Obviously, in prime power dimensions our methods can be
applied almost literally. When the dimension is not a prime power, and the whole set of Hadamard matrices is unknown, this method of quantum state characterization will still work: we need only $p^2-1$ equations (excluding the normalization condition) to determine all the parameters of the state (\ref{SD}).  These equations can be obtained by applying a set of appropriately chosen unitary transformations.  Although it would not generally be possible to establish an analytical procedure similar to that described in Sec. 3, a numerical solution to the resulting system of equation can still  be found.

Finally, we would like to mention a recent paper \cite{delgado} where a
tomographic reconstruction of a single qubit by applying methods of quantum
state discrimination was proposed.

This work is partially supported by the Grant 106525 of CONACyT (Mexico).

\end{document}